\documentclass[letterpaper]{jpconf}
\usepackage{graphicx}

\def\lamb#1#2{$^{#1}_{\Lambda}${#2}}
\def\lam#1#2{$^{#1}_{~\Lambda}${#2}}
\def\Kpi{($K^-,\pi^-$) }
\def\Kpist{($K^-_{stop},\pi^-$) }
\def\Kpig{($K^-,\pi^- \gamma$) }
\def\piK{($\pi^+,K^+$) }
\def\piKg{($\pi^+,K^+\gamma$) }

\begin{document}
\title{Structure of p-shell hypernuclei}

\author{D J Millener}

\address{Brookhaven National Laboratory, Upton, NY 11973, USA}

\ead{millener@bnl.gov}

\begin{abstract}
Shell-model calculations that include both $\Lambda$ and $\Sigma$ 
configurations with p-shell cores are used to interpret
$\gamma$-ray transitions in \lamb{7}{Li}, \lamb{9}{Be}, \lam{10}B, 
\lam{11}{B}, \lam{12}{C}, \lam{15}{N}, and \lam{16}{O} observed 
with the Hyperball array of Ge detectors. It is shown that the data 
puts strong constraints on the spin dependence of the $\Lambda N$ 
effective interaction and that the $\Lambda$-$\Sigma$ coupling
plays an important role.
\end{abstract}

\section{Introduction}
\label{sec:intro}

 Beginning in the 1970's, experiments at CERN, BNL, and KEK using
the \Kpist\ and in-flight \Kpi\ and \piK\ reactions (see table 2 
of \cite{hashimoto06}) established that the $\Lambda$ single-particle
energies in hypernuclei up to $^{208}_{\ \ \Lambda}$Pb form a
textbook example of single-particle behavior, describable by a 
Woods-Saxon well of depth $28-30$ MeV~\cite{hashimoto06,millener88} 
(consistent with earlier analyses of $B_\Lambda$ values from emulsion 
experiments~\cite{davis86}).

 The best energy resolution achieved in such experiments was
1.45 MeV using a thin carbon target~\cite{hotchi01}. Recently,
$(e,e'K^+)$ experiments in Hall A and Hall C at JLab have achieved
resolutions of $\sim 700$ keV~\cite{cusanno10} and $\sim 400$
keV~\cite{hashimoto10}, respectively. In general, this is still
insufficient to measure the doublet splittings that result when 
a $\Lambda$ in an $s$ orbit couples to a state of a p-shell nucleus 
with non-zero spin. 

 Because the $\Lambda$ is in an $s$ orbit, only a spatial monopole 
interaction operates between the nucleon and the $\Lambda$ in such
doublets. This  means that there is little configuration mixing between 
members of different doublets and that the doublet splittings depend 
almost entirely on the spin-spin, $\Lambda$ spin-orbit, and tensor 
components of the effective $\Lambda N$ interaction, together with a 
contribution from $\Lambda$-$\Sigma$ coupling. Given that 
data on the free $YN$ ($\Lambda N\!+\!\Sigma N$) interaction are very 
sparse and essentially constrain only spin-averaged s-wave scattering
for $\Lambda N$, the $s_\Lambda$ doublet spacings are an important source
of information on the $YN$ interaction. This requires hypernuclear
$\gamma$-ray spectroscopy. 

  The earliest measurements were made with NaI detectors 
\cite{bedjidian79,may83}. The excited $1^+$ states
of \lamb{4}{H} and \lamb{4}{He} were found to be at 1.04(4) MeV and
1.15(4) MeV, respectively~\cite{bedjidian79}. With just a $\Lambda N$
interaction, this implies that the spin-singlet central interaction
is more attractive than the triplet, as is also necessary to obtain
a $1/2^+$ hypertriton. Now, it is recognized that $\Lambda$-$\Sigma$
coupling contributes significantly to the $1^+\!-\!0^+$ spacing
mainly by increasing the binding energy of the $0^+$ ground state;
this also permits a consistent description of the binding energies
of the s-shell hypernuclei~\cite{akaishi00,hiyama01,nogga02,nemura02}.
The observation of $\gamma$-rays in \lamb{7}{Li} and 
\lamb{9}{Be}~\cite{may83} enabled progress to be made in the
use of shell-model calculations to extract information on the 
spin-dependence of the $\Lambda N$ interaction for p-shell 
hypernuclei~\cite{millener85} after the pioneering efforts of 
Gal, Soper, and Dalitz~\cite{gsd71} in which the ground-state 
$B_\Lambda$ values alone proved insufficient to give fits with any 
degree of uniqueness.
 
 However, the superior ($\sim$  keV) resolution 
of Ge detectors turns out to be essential. Following a pioneering experiment
at BNL~\cite{chrien90}, the capabilities of a large-acceptance Ge detector
array (Hyperball) have been exploited in a series of experiments
on p-shell targets carried out at KEK and BNL between 1998 and
2005 using the \piKg\ and \Kpig\ reactions, respectively (see table 3 of
\cite{hashimoto06}). 
As well as $\gamma$-ray transitions between bound states of the primary
hypernucleus, $\gamma$-ray transitions are often seen from
daughter hypernuclei formed by particle emission (most often a proton)
from unbound states of the primary hypernucleus.
 
\begin{center}
\begin{figure}[p]
\includegraphics*[width=15.0cm]{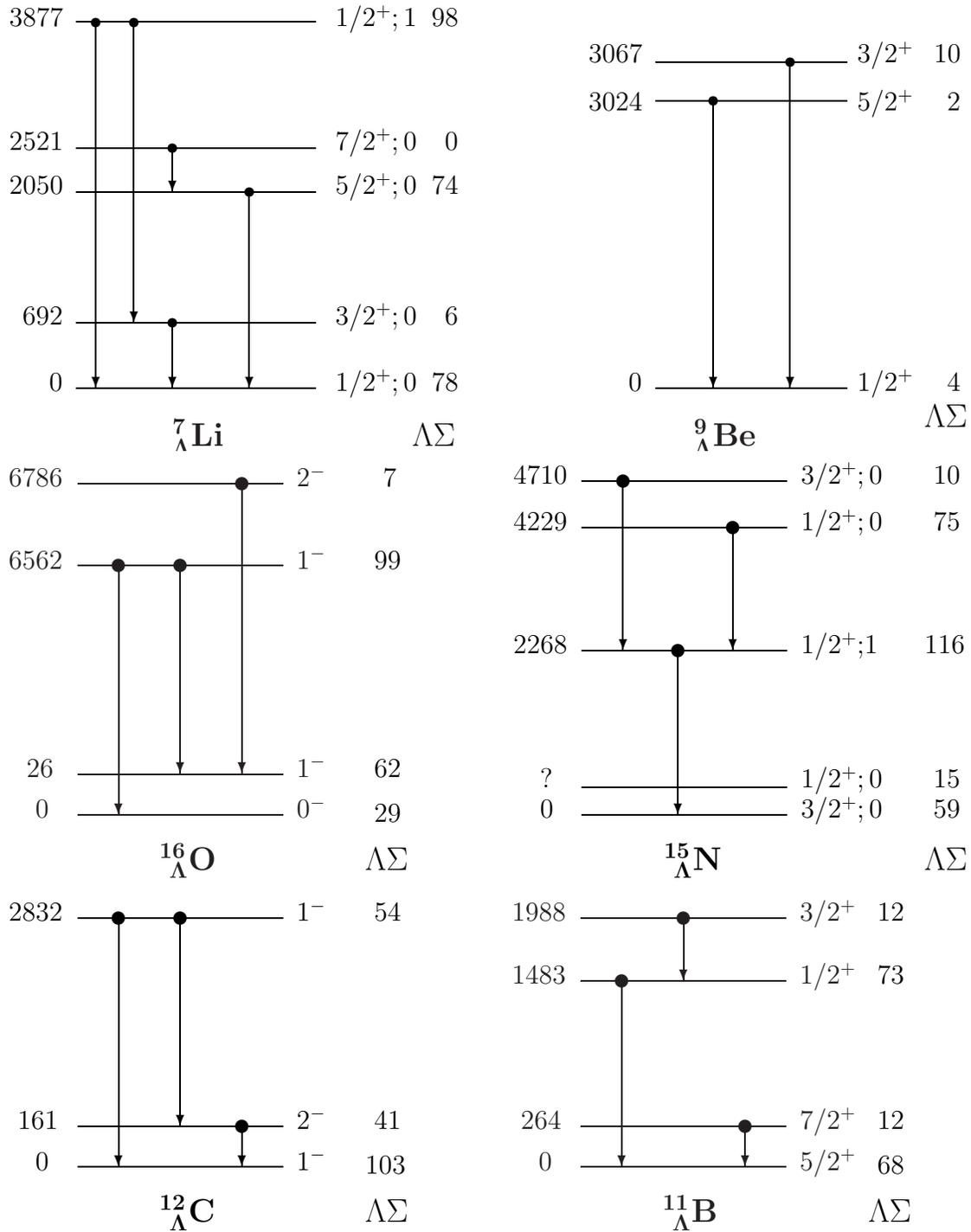}
\caption{The spectra of \lamb{7}{Li}, \lamb{9}{Be}, \lam{16}{O}, and
\lam{15}{N}, \lam{12}{C}, and \lam{11}{B} determined from experiments 
KEK E419, E518, E566, and BNL E930 with 
the Hyperball detector. All energies are in keV. The arrows denote 
observed $\gamma$-ray transitions. For each state the calculated 
energy shifts due to $\Lambda$-$\Sigma$ coupling are given.}
\label{fig:p-shell}
\end{figure}
\end{center}

 A total of 22 $\gamma$-rays, including nine doublet spacings, have 
now been observed with the Hyperball
or Hyperball-2 Ge-detector arrays \cite{hashimoto06,ukai08} in six p-shell 
hypernuclei, namely \lamb{7}{Li}, \lamb{9}{Be}, \lam{11}{B}, \lam{12}{C}, 
\lam{15}{N}, and \lam{16}{O}. The newest information concerns three 
$\gamma$-ray transitions each in \lam{12}{C} and \lam{11}{B} from 
KEK E566~\cite{ma10,tamura10} that used the \piKg\ reaction on a 
$^{12}$C target and the Hyperball-2 detector. The placements of 
most these $\gamma$ rays are summarized in figure~\ref{fig:p-shell}
(except for some transitions involving higher levels of \lam{11}{B}).

 After a brief description of the shell-model calculations and the
parametrization of the $YN$ interactions, a $\Lambda N$ parameter
set that fits \lamb{7}{Li} and another that fits the heavier
p-shell hypernuclei are given in section~\ref{sec:shell}. The fitting
procedure is described in section~\ref{sec:p-shell}. The details of 
these fits have been covered extensively in recent publications
\cite{millener07,millener08,millener09, millener10}.  
Section~\ref{sec:bb} gives matrix elements derived from
free $YN$ interactions and conclusions are presented in
section~\ref{sec:summary}.

\section{Shell-model calculations}
\label{sec:shell}

 Shell-model calculations for p-shell hypernuclei start with the
Hamiltonian 
\begin{equation}
 H = H_N + H_Y + V_{NY} \; ,
\label{eq:hamyn}
\end{equation}
where $H_N$ is an empirical Hamiltonian for the p-shell core,
the single-particle $H_Y$ supplies the $\sim 80$\,MeV mass difference
between $\Lambda$ and $\Sigma$, and $V_{NY}$ is the $YN$ interaction.
The shell-model basis states are chosen to be of the form
$|(p^n\alpha_{c}J_{c}T_{c},j_Yt_Y)JT\rangle$,
where the hyperon is coupled in angular momentum and isospin
to eigenstates of the p-shell Hamiltonian for the core, with up to three
values of $T_c$ contributing for $\Sigma$-hypernuclear states. This
is known as a weak-coupling basis and, indeed, the mixing of
basis states in the hypernuclear eigenstates is generally
very small. In this basis, the core energies are taken from
experiment where possible and from the p-shell calculation otherwise.

 To perform shell-model calculations, the $YN$ interaction
is written as \cite{auerbach83}
\begin{equation}
 V  =  \sum_\alpha C(\alpha)  \left[\left[ a^{+}_{j_N}\widetilde{a}_{j_N'}
\right]^{J_\alpha T_\alpha}\left[ a^{+}_{j_Y}
\widetilde{a}_{j_Y'}\right]^{J_\alpha T_\alpha}\right]^{00} \; ,
\label{eq:vcross}
\end{equation}
where $C(\alpha)$ represents linear combinations of the two-body
matrix elements, $\alpha$ stands for all the quantum numbers, and the 
tilde denotes properly phased annihilation operators. 
The basic input from the p-shell calculation is then a set of 
one-body density-matrix elements (OBDME) between all pairs of nuclear
core states that are to be included in the hypernuclear shell-model 
calculation. From the isospins of the hyperons, 
it is clear that only isoscalar OBDME are needed for coupling $\Lambda$ 
configurations while isovector OBDME are needed for coupling $\Lambda$ 
configurations to $\Sigma$ configurations. For the p-shell wave
functions of the core, early hypernuclear shell-model calculations
of the $\gamma$-ray era~\cite{millener85,fetisov91} used one or more
of the three Cohen and Kurath interactions~\cite{cohen65}. More recently,
interactions fitted to p-shell energy-level data with the strength of 
the tensor interaction fixed to reproduce the cancellation in the $^{14}$C
$\beta$ decay ($^{14}$N ground-state wave function) have been
used~\cite{millener07}.  

 The $\Lambda N$ effective interaction can be  
written~\cite{gsd71}
\begin{equation}
V_{\Lambda N}(r)  =  V_0(r) + V_{\sigma}(r)~ \vec{s}_N\cdot
 \vec{s}_{\Lambda} +  V_{\Lambda }(r)~\vec{l}_{N\Lambda }\cdot
\vec{s}_{\Lambda}  + V_{\rm N}(r)~\vec{l}_{N \Lambda }\cdot
\vec{s}_{N} +  V_{\rm T}(r)~S_{12}\; ,
\label{eq:vlam}
\end{equation}
where $S_{12} = 3(\vec{\sigma}_{N}\cdot\vec{r})(\vec{\sigma}_{\Lambda}
\cdot\vec{r})-\vec{\sigma}_{N}\cdot\vec{\sigma}_{\Lambda}$. The five
$p_N s_\Lambda$ two-body matrix elements depend on the radial integrals 
associated with each component in equation~\ref{eq:vlam}. They are denoted 
by the parameters $\overline{V}$, $\Delta$, $S_\Lambda$, $S_N$ and 
$T$~\cite{gsd71}. By convention~\cite{gsd71}, $S_\Lambda$ and $S_N$ are
actually the coefficients of $\vec{l}_N\cdot\vec{s}_\Lambda$ and
$\vec{l}_N\cdot\vec{s}_N$. Then, the operators associated with
$\Delta$ and $S_\Lambda$ are $\vec{S}_N\cdot \vec{s}_{\Lambda}$
and $\vec{L}_{N}\cdot \vec{s}_{\Lambda}$. In an LS basis for the core,
the matrix elements of $\vec{S}_N\cdot \vec{s}_{\Lambda}$ are diagonal
(similarly for $\vec{L}_{N}\cdot \vec{s}_{\Lambda} = 
(\vec{J}_N -\vec{S}_N)\cdot \vec{s}_{\Lambda}$) and therefore depend
only on the intensities of the different $L_c$ and $S_c$ in the
core wave functions. Because supermultiplet symmetry $[f_c]K_cL_cS_cJ_cT_c$ 
is a rather good symmetry for p-shell core states, only a few values 
of $L_c$ and $S_c$ are important leading to a simple understanding
of the contributions from $\Delta$ and $S_\Lambda$.
$\overline{V}$ contributes only to the overall binding energy;
$S_N$ does not contribute to doublet splittings in the weak-coupling
limit but augments the nuclear spin-orbit interaction and contributes
to the spacings between states based on different core states; in general,
there are not simple expressions for the coefficients of $T$.

 The parametrization of equation~\ref{eq:vlam} applies to the direct 
$\Lambda N$ interaction, the $\Lambda N$--$\Sigma N$ coupling interaction, 
and the direct $\Sigma N$ interaction for both isospin 1/2 and 3/2.
A set of parameters that fits the energy spacings between levels of 
\lamb{7}{Li} and \lamb{9}{Be} is (parameters in MeV)
\begin{equation}
\Delta= 0.430\quad S_\Lambda =-0.015\quad {S}_{N}
 = -0.390 \quad {T}=0.030 \; ,
\label{eq:param7}
\end{equation}
while a set that fits the heavier p-shell hypernuclei is
\begin{equation}
\Delta= 0.330\quad S_\Lambda =-0.015\quad {S}_{N}
 = -0.350 \quad {T}=0.0239 \; .
\label{eq:param11}
\end{equation}
The corresponding matrix 
elements for the $\Lambda$-$\Sigma$ coupling interaction, based on the 
G-matrix calculations of \cite{akaishi00} for the nsc97$e,f$
interactions \cite{rijken99}, are \cite{millener08,millener07}
\begin{equation}
\overline{V}' = 1.45\quad \Delta'= 3.04\quad S_\Lambda' = S_N' = -0.09
\quad T' = 0.16 \; .
\label{eq:paramls}
\end{equation}
These parameters are kept fixed in the present calculations.

 From equation~\ref{eq:paramls} it is clear that the central terms,
$\overline{V}'$ and $\Delta'$, of $\Lambda$--$\Sigma$ coupling 
interaction are important. Formally, one could include an overall
factor $t_N\cdot t_{\Lambda\Sigma}$ in the analog of equation~\ref{eq:vlam}
that defines the interaction, where $t_{\Lambda\Sigma}$ is the
operator that converts a $\Lambda$ into a $\Sigma$. Then, the
core operator associated with  $\overline{V}'$ is $T_N = \sum_i t_{Ni}$.
This leads to a non-zero matrix element only between $\Lambda$ and
$\Sigma$ states that have the same core, with the value
\begin{equation}
 \langle (J_cT,s_\Sigma)JT |V'_{\Lambda\Sigma}|(J_cT,s_\Lambda)JT\rangle
 = \sqrt{4/3}\ \sqrt{T(T+1)}\ \overline{V}'\; ,
\label{eq:fermi}
\end{equation}
in analogy to Fermi $\beta$ decay of the core nucleus. Similarly,
the spin-spin term involves $\sum_i s_{Ni}t_{Ni}$ for the core and 
connects core states that have large Gamow-Teller matrix elements 
between them. These are states with mainly the same $[f_c]K_cL_c$
because the Gamow-Teller operator can't change any spatial quantum 
numbers. For a $T\!=\!0$ core nucleus, only $\Delta'$ contributes.

\begin{center}
\begin{table}
\caption{Doublet spacings in p-shell hypernuclei. Entries in the top 
(bottom) half of the table are calculated using the parameters in 
equation~\ref{eq:param7} (equation~\ref{eq:param11}). The individual 
contributions do not sum to exactly $\Delta E^{th}$, which comes
from the diagonalization, because small contributions from the
energies of admixed core states are not included. The coefficients
of  $\Delta$, $S_\Lambda$, $S_N$, and $T$ can be obtained by dividing
the contributions in the table by the values of the parameters in 
equation~\ref{eq:param7} or equation~\ref{eq:param11}. Alternatively, they 
are given in Refs.~\cite{millener07,millener08,millener09, millener10}.}
\label{tab:spacings}
\begin{tabular*}{\textwidth}{@{}l@{\extracolsep{\fill}}ccrrrrrrr}
\br
 & $J^\pi_u$ & $J^\pi_l$ & $\Lambda\Sigma$ & $\Delta$ & $S_\Lambda$ & $S_N$ 
 & $T$ & $\Delta E^{th}$ & $\Delta E^{exp}$  \\
\mr
\lamb{7}{Li} & $3/2^+$ & $1/2^+$ & 72 & 628 & $-1$ & 
$-4$ & $-9$ &  693 & 692 \vspace{1pt}\\
\lamb{7}{Li} & $7/2^+$ & $5/2^+$ & 74 & 557 & $-32$ & 
$-8$ & $-71$ &  494 & 471 \vspace{1pt}\\
\lamb{8}{Li} & $2^-$ & $1^-$ & 151 & 396 & $-14$ & 
$-16$ & $-24$ &  450 & (442) \vspace{1pt}\\
\lamb{9}{Li} & $5/2^+$ & $3/2^+$ & $116$ & $530$ & 
$-17$ & $-18$ & $-1$ &  $589$ & \vspace{1pt}\\
\lamb{9}{Li} & $3/2^+_2$ & $1/2^+$ & $-80$ & $231$ & 
$-13$ & $-13$ & $-93$ &  $-9$ & \vspace{1pt}\\
\lamb{9}{Be} & $3/2^+$ & $5/2^+$ & $-8$ & $-14$ & 
$37$ & $0$ & $28$ &  $44$ & 43 \vspace{5pt}\\
\lam{10}{B} & $2^-$ & $1^-$ & $-15$ & 188 & $-21$ & 
$-3$ & $-26$ &  120 & $<100$ \vspace{1pt}\\
\lam{11}{B} & $7/2^+$ & $5/2^+$ & 56 & 339 & $-37$ & 
$-10$ & $-80$ &  267 & 264 \vspace{1pt}\\
\lam{11}{B} & $3/2^+$ & $1/2^+$ & 61 & 424 & $-3$ & 
$-44$ & $-10$ &  475 & 505 \vspace{1pt}\\
\lam{12}{C} & $2^-$ & $1^-$ & 61 & 175 & $-12$ & 
$-13$ & $-42$ &  153 & 161 \vspace{1pt}\\
\lam{15}{N} & $1/2^+_1$ & $3/2^+_1$ & 44 & 244 & 34 & 
$-8$ & $-214$ &  99 &  \vspace{1pt}\\
\lam{15}{N} & $3/2^+_2$ & $1/2^+_2$ & 65 & 451 & $-2$ & 
$-16$ & $-10$ &  507 & 481 \vspace{1pt}\\
\lam{16}{O} & $1^-$ & $0^-$ & $-33$ & $-123$ & $-20$ & 
1 & 188 &  23 & 26 \vspace{1pt}\\
\lam{16}{O} & $2^-$ & $1^-_2$ & 92 & 207 & $-21$ & 
1 & $-41$ &  248 & 224 \vspace{1pt}\\
\br
\end{tabular*}
\end{table}
\end{center}

 Perturbatively, the energy shift that $\Lambda$-hypernuclear state
gets from mixing with a $\Sigma$-hypernuclear state goes as the 
mixing matrix element squared over an energy denominator that is of
the order of 80\,MeV. Equation~\ref{eq:fermi} shows that the contribution
from $\overline{V}'$ increases quadratically with the isospin for
neutron-rich hypernuclei. The total Gamow-Teller strength, proportional
to $N-Z$, increases linearly with isospin. However, not all the 
Gamow-Teller strength is operative for a given $J$. An important
feature is the magnitude and sign of the contribution from $\Delta'$
to the ``diagonal'' matrix element in equation~\ref{eq:fermi}. This is
illustrated by the off-diagonal matrix elements $v(J)$ in a
simple $s^3s_\Lambda$ plus $s^3s_\Sigma$ model \cite{akaishi00,millener07}
for the $0^+$ and $1^+$ states of \lamb{4}{He} (or \lamb{4}{H})
where
\begin{equation}
 v(0) = \overline{V_s}' + 3/4\Delta_s'\:,\quad  
 v(1) = \overline{V_s}' - 1/4\Delta_s' \; .
\label{eq:s-shell}
\end{equation}
The pure s-state matrix elements take roughly double the p-shell
values in equation~\ref{eq:fermi} leading to 7.46 MeV for $v(0)$ and a 
much smaller value for $v(1)$. The resultant energy shift is $\sim 700$ 
keV for the \lamb{4}{He} ground state. Thus, the $\Lambda N$ 
spin-spin interaction and $\Lambda$-$\Sigma$ coupling make 
comparable contributions to the $1^+\!-\!0^+$ doublet splitting.
Few-body calculations have confirmed this 
basic picture \cite{hiyama01,nogga02,nemura02}.

\section{Analysis of the p-shell hypernuclei}
\label{sec:p-shell}

 Table~\ref{tab:spacings} gives the breakdown of the contributions
from $\Lambda$-$\Sigma$ coupling and the $\Lambda N$ interaction
parameters to all 9 of the measured doublet spacings and several more
doublets spacings of interest in \lamb{8}{Li}, \lamb{9}{Li}
\lam{10}{B}, and \lam{15}{N}. The strategy for fixing the values
of the parameters goes as follows.

 The \lamb{9}{Be} doublet spacing demands a small value for $S_\Lambda$. 
This was already clear from the limit of 100 keV placed on the spacing
using NaI detectors~\cite{may83,millener85}. This limit relies on theory 
only to the extent that the doublet members are expected to be 
populated almost equally in the \Kpi\ reaction~\cite{dalitz78}.
The unbound $2^+$ excited state of $^8$Be (bound in the presence
of the $\Lambda$) has dominantly $L\!=\!2$ and $S\!=\!0$, in which
case the $3/2^+$ lies above the $5/2^+$ state by $-5/2 S_\Lambda$.
The small $S\!=\!1$ admixtures in the $2^+$ wave function, necessary
to account for the Gamow-Teller decays of $^8$Li and $^8$B, lead
to small contributions from $\Delta$ and $T$ to the doublet spacing.
The small Gamow-Teller matrix elements also mean that the 
contribution from $\Lambda$-$\Sigma$ coupling ($\Delta'$) is small
and it so happens that the contributions other than from $S_\Lambda$
more or less cancel.  The parameter set chosen puts the $3/2^+$ state 
above the $5/2^+$ state but the order is not determined in the 
original experiment~\cite{akikawa02}. However, in the 2001 run of 
BNL E930 on a $^{10}$B target, only the upper level is seen strongly 
following proton emission from \lam{10}{B}~\cite{tamura05}. 
It can then be deduced that the $3/2^+$ state is the upper member of the 
doublet~\cite{millener07}.

 Four of the five $\gamma$ rays in \lamb{7}{Li} were observed
in the first Hyperball experiment~\cite{tamura00} while the
excited-state doublet spacing was observed following $^3$He
emission from $0s$-hole states in \lam{10}{B}~\cite{ukai06}.
The ground state of $^6$Li is mainly $L\!=\!0$ and 
$S\!=\!1$~\cite{millener07} which means that the
spins ground-state doublet members in \lamb{7}{Li} are
all due to intrinsic spin and the doublet spacing is given 
by $3/2\;\Delta$ plus the contribution from $\Lambda$-$\Sigma$
coupling (due to $\Delta'$). The latter accounts for $\sim\!10$\%
of the spacing. The excited $3^+$ state of $^6$Li is purely
$L\!=\!2$, $S\!=\!1$ leading, in this limit, to a doublet
spacing of~\cite{dalitz78}
\begin{equation}
\Delta E = 7/6\ \Delta + 7/3\ {S}_\Lambda  - 14/5\ T \; .
\label{eq:75doublet}
\end{equation}
Again, the spin-spin interaction dominates but the spacing is
reduced by contributions from $S_\Lambda$ and $T$. Initially,
$T$ was taken to be small based on the available $YN$ 
interactions~\cite{millener05}. The first indication of a
substantial negative value for $S_N$ came from the excitation
energy of the $5/2^+$ state~\cite{may83,millener05,fetisov91} 
which, being based on the lowest member of an $L\!=\!2$, $S\!=\!1$ 
triplet, is lowered by an enhanced nuclear spin-orbit interaction
(see table~\ref{tab:excitation}); $S_N$ also makes an important 
contribution to the excitation energy of the $1/2^+;1$ state.

 The $A\!=\!7$ hypernuclei have also been extensively studied using
three-body~\cite{hiyama96,hiyama99} and four-body~\cite{hiyama06,hiyama09}
models. These models lack a tensor interaction and explicit 
$\Lambda$-$\Sigma$ coupling but can treat the radial structure of
these hypernuclei. Indeed, a reduced B(E2) for the $5/2^+\to 1/2^+$ 
transition relative to the known $3^+\to 1^+$ core transition in $^6$Li 
was predicted due to a contraction brought about by the extra binding 
energy in the presence of a $\Lambda$, as subsequently found 
experimentally~\cite{tanida01}.

 The determination of $T$ rests on the measurement of the ground-state
doublet spacing in \lam{16}{O}~\cite{ukai04,ukai08}. Because the $1^-$,
$0^-$ doublet spacing, given by
\begin{equation}
\Delta E = -1/3\ \Delta + 4/3\ {S}_\Lambda  + 8\ T \; ,
\label{eq:l16o}
\end{equation}
could be small, the experiment was set up to measure the $\gamma$-ray
energies from the $1^-_2$ state to the members of the ground-state
doublet. A weak transition, tentatively assigned as from the $2^-$
level, was also observed~\cite{ukai08}, thus providing the $2^-$, 
$1^-$ excited-state doublet spacing.

\begin{center}
\begin{table}
\caption{Non-doublet excitation energies of states in p-shell hypernuclei. 
The entry for \lamb{7}{Li} is calculated using the parameters in 
equation~\ref{eq:param7}. The remainder are calculated using 
equation~\ref{eq:param11}. See table~\ref{tab:spacings} for more details.
$\Delta E_c$ is the excitation energy of the core state. Taking the
centroid of doublets would eliminate most of the dependence on
$\Delta$, $S_\Lambda$, and $T$. All energies are in keV.}
\label{tab:excitation}
\begin{tabular*}{\textwidth}{@{}l@{\extracolsep{\fill}}ccrrrrrrr}
\br
 & $J^\pi;T$ & $\Delta E_c$ & $\Lambda\Sigma$ & $\Delta$ & $S_\Lambda$ & $S_N$ 
 & $T$ & $E^{th}_x$ & $E^{exp}_x$  \\
\mr
\lamb{7}{Li} & $5/2^+$ & 2186 & 4 & 77 & 17 & 
$-288$ & 33 &  2047 & 2050 \vspace{1pt}\\
\lamb{7}{Li} & $1/2^+;1$ & 3565 & $-23$ & 418 & 0 & 
$-82$ & $-3$ &  3883 & 3877 \vspace{1pt}\\
\lam{11}{B} & $1/2^+;0$ & 718 & 5 & $-88$ & $-19$ & 
391 & $-38$ &  968 & 1483 \vspace{1pt}\\
\lam{12}{C} & $1^-;1/2$ & 2000 & 49 & 117 & $-17$ & 
309 & 20 &  2430 & 2832 \vspace{1pt}\\
\lam{13}{C} & $3/2^+;0$ & 4439 & 1 & $-11$ & 22 & 
203 & $-22$ & 4630 & 4880 \vspace{1pt}\\
\lam{15}{N} & $1/2^+;1$ & 2313 & $-57$ & 86 & 11 & 
$-6$ & $-71$ &  2274 & 2268 \vspace{1pt}\\
\lam{15}{N} & $1/2^+_2;0$ & 3948 & $-16$ & $-208$ & 13 & 
473 & $-67$ & 4120 & 4229 \vspace{1pt}\\
\lam{16}{O} & $1^-_2;1/2$ & 6176 & $-70$ & $-207$ & $-2$ & 
524 & 170 &  6582 & 6562 \vspace{1pt}\\
\br
\end{tabular*}
\end{table}
\end{center}

 In the same experiment~\cite{ukai08}, three $\gamma$-ray transitions
in \lam{15}{N} were observed following proton emission from unbound
states of \lam{16}{O}. In the simplest model of $p_{1/2}^{-2}s_\Lambda$, 
the ground-state doublet spacing should be just $3/2$ that of
\lam{16}{O}~\cite{dalitz78,millener05} with the $1/2^+$ state lowest
and should be measureable from the decays of the 2268-keV $1/2^+;1$ 
level (see figure~\ref{fig:p-shell}). This would provide a check on 
the value of $T$ from \lam{16}{O}. However, only one very sharp 
$\gamma$ ray corresponding to a long lifetime is observed. In addition, 
the lifetime of the 2268-keV level is measured to be 15 times longer
than that of the $0^+;1$ core state~\cite{ukai08}. The core
transition is a weak, mainly orbital, M1 transition because
the spin matrix element is almost zero (cf. $^{14}$C $\beta^-$ decay).
It turns out that small $1^+_2;0\times s_\Lambda$ admixtures
introduce a strong spin M1 matrix element that produces 
strong cancellations, the more so for the decay to the $1/2^+$
member of the ground-state doublet~\cite{millener07,ukai08}.
The ground-state is predicted to be $3/2^+$, opposite to that expected 
from the simple model mentioned above, and this is confirmed 
by recent results on the mesonic weak decay of 
\lamb{15}{N}~\cite{agnello09,gal09}.
The other two observed $\gamma$ rays determine the spacing of the
excited-state doublet. Here the core state is largely $L\!=\!0$,
$S\!=\!1$ and, as for the ground-state doublet in \lamb{7}{Li}, 
the spacing is driven by $\Delta$ and is well fitted by the 
value in equation~\ref{eq:param11}. The excited-state doublet spacing
in \lam{11}{B} provides a similar example.
 
 The data on \lam{12}{C} and \lam{11}{B} in figure~\ref{fig:p-shell}
come from KEK E566~\cite{ma10,tamura10}, the \lam{11}{B}
$\gamma$ rays following proton emission from unbound states of
\lam{12}{C} (a total of six $\gamma$ rays are known in 
\lam{11}{B}~\cite{miura05,hashimoto06}).
 
  As can be seen from table~\ref{tab:spacings}, there is a
consistent description of the doublet spacings once a larger
value of $\Delta$ is taken for \lamb{7}{Li}. The ground-state
doublet of \lamb{8}{Li} is included because there is a candidate
$\gamma$-ray \cite{chrien90}. The ground-state
doublet of \lamb{9}{Li} could be measured using the 
$^9$Be$(e,e'K^+)$\lamb{9}{Li} reaction at Jefferson Laboratory
(the spacing is 470 keV with the other parameter set). The
excited-state doublet of \lamb{9}{Li} is included to show that
the contributions from $\Delta$ and $\Lambda$-$\Sigma$ coupling 
are not always of the same sign.

 The remaining point with regard to doublet spacings concerns
the ground-state doublet spacings of \lam{12}{C} and \lam{10}{B}.
The limit on the \lam{10}{B} spacing~\cite{chrien90} was
the reason that Fetisov et al. chose a small value of 
$\Delta$~\cite{fetisov91}. With the inclusion of $\Lambda$-$\Sigma$ 
coupling, it can be seen from table~\ref{tab:spacings} that 
$\Lambda$-$\Sigma$ coupling reduces the spacing in \lam{10}{B}
and increases it in \lam{12}{C}~\cite{millener07,millener10}. 
The recent measurement of the ground-state doublet spacing in 
\lam{12}{C} provides an important check on this effect. It is 
possible to reduce the \lam{10}{B} spacing further by 
adjustments to the $\Lambda$-$\Sigma$ coupling parameters in 
equation~\ref{eq:paramls}~\cite{millener10}. It should also be borne 
in mind that the protons are relatively loosely bound in \lam{10}{B}
($^9$B is unbound to proton emission) but the effect of this is hard
to estimate because the p-shell parentage is widely spread.

 Table~\ref{tab:excitation} shows the breakdown of contributions
to non-doublet excitation energies for the hypernuclei in 
figure~\ref{fig:p-shell}. This shows the influence of $S_N$ on
these spacings; by taking the centroid of doublets one could
remove most of the contributions from $\Delta$, $S_\Lambda$,
and $T$ but the contribution from $S_N$ would not be much
affected, being essentially the same for doublet members
(table~\ref{tab:spacings}). As can be seen $S_N$ always
contributes in the right direction to improve agreement
with experiment, but leads to an underestimate of the
excitation energies for \lam{11}{B}, \lam{12}{C}, and
\lam{13}{C} near the middle of the shell (there is sensitivity
to the core wave functions). Possible reasons for this
discrepancy will be addressed in section~\ref{sec:summary}.

\section{Matrix elements from baryon-baryon interactions}
\label{sec:bb}

  In the preceding section, the $\Lambda N$ matrix elements have
been treated as parameters independent of mass number, except
(essentially) for the larger value of $\Delta$ for \lamb{7}{Li}. 
This is reasonable because the stable p-shell nuclei exhibit almost 
constant rms charge radii~\cite{millener07}. To illustrate this 
point for hypernuclei, we show in table~\ref{tab:woods} the values 
obtained for the parameters when they are calculated from fixed 
radial forms of the potentials using Woods-Saxon wave functions. 
Of the nsc97 interactions \cite{rijken99}, only nsc97f gives a
 value for $\Delta$ that comes close to the empirical value. 
The radial representation of nsc97f then forms a starting point 
for scaling the strengths in the various central, spin-orbit, 
antisymmetric, and tensor channels to obtain a fit to either of 
the empirically determined parameter sets; fit-djm is the result 
for the \lamb{7}{Li} set. Most recent $YN$ models (such as 
esc04~\cite{rijken06}, esc08~\cite{rijken10},
or the new J\"{u}lich models~\cite{haidenbauer07}), use some
constraint to obtain a more attractive singlet s-wave
interaction than triplet to bind the hypertriton, and give a 
reasonable $A\!=\!4$ $0^+$/$1^+$ doublet spacing.
The Yukawa representation of the esc04a potential
comes from the work of Halderson~\cite{halderson08} (similarly
esc08a).

\begin{center}
\begin{table}
\caption{Parameter values for p-shell and s-shell hypernuclei
calculated using Woods-Saxon wave functions and Gaussian or Yukawa 
representations of $\Lambda N$-$\Lambda N$ G-matrix elements.}
\label{tab:woods}
\begin{tabular*}{\textwidth}{@{}l@{\extracolsep{\fill}}cccccccc}
\br
 & & \multicolumn{5}{c}{p-shell} & \multicolumn{2}{c}{s-shell}\\
 &  & $\overline{V}$ & $\Delta$ & $S_\Lambda$ & $S_N$ & $T$ & $\overline{V}_s$ 
& $\Delta_s$  \\
\mr
 fit-djm & \lamb{7}{Li} & $-1.142$ & 0.438 & $-0.008$ & $-0.414$ & 0.031 & 
$-1.387$ & 0.497 \vspace{1pt} \\
 fit-djm & \lam{16}{O} & $-1.161$ & 0.441 & $-0.007$ & $-0.401$ & 0.030 & 
 & \vspace{1pt}\\
 nsc97f & \lamb{7}{Li} & $-1.086$ & 0.421 & $-0.149$ & $-0.238$ & 0.055 & 
$-1.725$ & 0.775 \vspace{1pt}\\
  esc04a & \lamb{7}{Li} & $-1.287$ & 0.381 & $-0.108$ & $-0.236$ & 0.013 & 
$-1.577$ & 0.850 \vspace{1pt}\\
  esc08a & \lamb{7}{Li} & $-1.221$ & 0.146 & $-0.074$ & $-0.241$ & 0.055 & 
$-1.796$ & 0.650 \vspace{1pt}\\
\br
\end{tabular*}
\end{table}
\end{center}

 The fit-djm results for \lamb{7}{Li} and \lam{16}{O} are calculated
from the same interaction for Woods-Saxon radii that scale as $A^{1/3}$
and depths that are fitted to the experimental binding energies.
As can be seen from the first two lines of table~\ref{tab:woods}, 
the calculated parameters stay remarkably constant (the
rms radii of the p nucleon orbits stay very mearly constant at around
2.9 fm while the rms radius of the $s_\Lambda$ orbit goes from
2.34 fm for \lam{16}{O} to 2.60 fm for \lamb{7}{Li}). The
cancellation between the symmetric and antisymmetric spin-orbit
interactions for nsc97f, esc04a, and esc08a is in the right direction,
but not large enough, to reproduce the very small empirical value
of $S_\Lambda$.

In the p shell, $\Delta$ receives contributions from the 
spin-dependence in both relative s states and p states
\cite{millener07} (cf. \cite{halderson08}). These the p-wave
contributions are repulsive for both spin channels in nsc97f, repulsive 
for singlet in esc04a and esc08a, and attractive for both in fit-djm. 
Despite the smaller size of p-wave matrix elements relative to s-wave
matrix elements, the contribution to $\Delta$ can be substantial
when the singlet and triplet p-wave interactions are of opposite
sign, as they are for esc04a and esc08a. This results in an
unsatisfactorily small value of $\Delta$ for esc08a. An effect
of this phenomenon is also seen in the substantial variation in the 
s-shell parameters (similarly calculated) in table~\ref{tab:woods} where
the contributions from odd-state central interactions are absent. 
A few-body calculation of 
\lamb{4}{H} and \lamb{4}{He} versus \lamb{7}{Li} would then test
the even/odd character of the central interaction in addition to the
role of $\Lambda$-$\Sigma$ coupling \cite{nogga02}.

 Finally, table~\ref{tab:lamsig} shows calculated $\Lambda$-$\Sigma$ 
coupling parameter values for several of the Nijmegen baryon-baryon 
interactions. These are dominantly effective central interactions that 
arise from the strong tensor interaction of $\Lambda$-$\Sigma$ 
coupling acting in second order. The esc04 interactions seem to have an
unphysical radial behavior \cite{halderson08}.

\begin{center}
\begin{table}
\caption{ $p_Ns_Y$ $\Lambda$-$\Sigma$ coupling parameters from several
of the Nijmegen baryon-baryon potentials.} 
\label{tab:lamsig}
\begin{tabular*}{\textwidth}{@{}l@{\extracolsep{\fill}}lrrrrr}
\br
Source &  Interaction  & $\bar{V}'$ & $\Delta'$ & $S_\Lambda'$ & 
$S_{N}'$ & $T'$ \\
\mr
Akaishi (s-shell) & NSC97e/f &  1.45  & 3.04 & $-0.09$ & $-0.09$ & 0.16  \\
Yamamoto & NSC97f &  0.96  & 3.62 & $-0.07$ & $-0.07$ & 0.31  \\
Halderson & NSC97e &  0.75  & 3.51 & $-0.45$ & $-0.24$ & 0.31  \\
Halderson & NSC97f &  1.10  & 3.73 & $-0.45$ & $-0.23$ & 0.30  \\
Halderson & ESC04a &  $-2.30$  & $-2.59$ & $-0.17$ & $-0.17$ & 0.23  \\
Halderson & ESC08a &  1.05  & 4.71 & $-0.07$ & $0.02$ & 0.32  \\
\br
\end{tabular*}
\end{table}
\end{center}

\section{Discussion}
\label{sec:summary}

  Section~\ref{sec:p-shell} and table~\ref{tab:spacings} show that the
doublet spacings in p-shell hypernuclei are rather well accounted for by
a consistent set of $\Lambda N$ interaction parameters $\Delta$, $S_\Lambda$,
and $T$ together with the contributions from $\Lambda$-$\Sigma$ coupling.
The main problem with consistency is that a larger value of the spin-spin
matrix element $\Delta$ is required near the beginning of the shell,
most certainly for \lamb{7}{Li}. The discussion in section~\ref{sec:bb}
shows that this is unlikely to be simply an effect of nuclear size.
Another possiblity is that restricting the core wave functions to
p-shell configurations is inadequate. When the core bases are expanded 
by adding higher configurations, the most important admixtures at the
beginning of the p shell involve nucleons excited from the s shell to
the p shell, For the specific case of $^6$Li, promoting an $np$ pair
from the s shell to the p shell forms a particularly stable
$\alpha$-like $2n2p$ system ($^8$Be) in the p shell in analogy to the 
classic low-lying $4p2h$ states in the $A\!=\!18$ nuclei. This
leaves an active $np$ pair in the s shell for which the $s_Ns_\Lambda$
matrix element is larger than the $p_Ns_\Lambda$ (central) matrix
element by roughly a factor of two. In contrast, beyond $^8$Be the
low-lying excitations involve nucleons excited from the p shell to
the sd shell. In $^{10}$B for example, the lowest $(sd)^2$ levels
are the 5.18-MeV $1^+;0$ level and the 7.56-MeV $0^+;1$ level. In
this case, the $(sd)_Ns_\Lambda$ matrix elements are smaller than
those for $p_Ns_\Lambda$.
 
 The other major discrepancy found in the present shell-model 
treatment is that the nuclear-spin-dependent spin-orbit term $S_N$ does 
not provide a large enough contribution for the mid-shell hypernuclei
(table~\ref{tab:excitation}). In the core nuclei, antisymmetric
spin-orbit interactions play an important role in the latter half of the
p-shell, mocking up the effect of $NNN$ interactions in producing
a $3^+$ ground state for $^{10}$B and enough ``spin-orbit'' splitting
at $A\!=\!15$. These can be obtained by averaging the $NNN$ interaction 
over the closed shell of $0s$ nucleons. For hypernuclei, the double 
one-pion exchange $\Lambda NN$ interaction \cite{gsd71} is independent 
of the $\Lambda$ spin and gives, when averaged over the $s_\Lambda$ 
wave function,  
\begin{equation}
\label{eq:lnn}
V_{NN}^{eff} = \sum_{klm} Q^k_{lm}(r_1,r_2) \left[\sigma_1,\sigma_2
\right]^k\cdot \left[C_l(\hat{r_1}),C_m(\hat{r_2})\right]^k\ \tau_1
\cdot\tau_2 \; .
\end{equation}
The $Q^0_{00}$ and  $Q^0_{22}$ terms give repulsive contributions to 
$B_\Lambda$ that depend quadratically on the number of p-shell nucleons 
in the core while $Q^1_{22}$ represents an anti-symmetric spin-orbit 
interaction that behaves rather like $S_N$ \cite{gsd71}, and could act
synergistically with the vector interactions in the core. This would 
help alleviate the general problem of failing to get enough spacing 
between states based on different core levels for hypernuclei near the 
middle of the p shell despite the fact that $S_N$ works in the right 
direction.

 Theoretically, then, the next step is to expand the basis for the nuclear 
core states and to explicitly include $\Lambda NN$ interactions. 
Experimentally, progress will come from experiments using the new 
Hyperball-J detector at J-PARC~\cite{tamura10}.

\ack

This work has been supported by the US Department of Energy under Contract
No. DE-AC02-98CH10886 with Brookhaven National Laboratory.

\section*{References}


\bibliographystyle{iopart-num}
\bibliography{inpc}

\end{document}